\documentclass{rnaastex}
\usepackage[utf8]{inputenc}
\usepackage{bm}
\usepackage{amsmath}
\usepackage{booktabs}
\usepackage{graphicx}

\shorttitle{TGAS Wide Binary Validation}
\shortauthors{Andrews et al.}

\begin{document}

\title{Validating TGAS wide binaries with {\it Gaia} DR2 Radial Velocities and Parallaxes}

\correspondingauthor{Jeff J.\ Andrews}
\email{andrews@physics.uoc.gr}

\author[0000-0001-5261-3923]{Jeff J.\ Andrews}
\affiliation{Foundation for Research and Technology-Hellas, 
100 Nikolaou Plastira St., 
71110 Heraklion, Crete, Greece}
\affiliation{Physics Department \& Institute of Theoretical \& Computational Physics, 
P.O.~Box 2208, 
71003 Heraklion, Crete, Greece}

\author{Julio Chanam\'e}
\affiliation{Instituto de Astrof\'isica, Pontificia Universidad Cat\'olica de Chile, Av.~Vicu\~na Mackenna 4860, 782-0436 Macul, Santiago, Chile}
\affiliation{Millennium Institute of Astrophysics, Santiago, Chile}

\author{Marcel A.~Ag\"ueros}
\affiliation{Department of Astronomy, Columbia University, 550 West 120th Street, New York, NY 10027, USA}

\section{}

Featuring greatly improved astrometry and precise radial velocities (RVs), the second data release (DR2) from {\it Gaia} \citep{2018arXiv180409365G} affords us a unique opportunity to test the validity of samples of stellar wide binaries. In \citet{2017MNRAS.472..675A}, we presented a set of wide binaries identified in the joint Tycho-{\it Gaia} Astrometric Solution \citep[TGAS;][]{2016A&A...595A...4L} catalog. Here, we use DR2 to confirm the low contamination rate of this sample, thereby verifying the effectiveness of our algorithm for application to DR2 data. 

We selected our binaries in \citet{2017MNRAS.472..675A} by matching data for the five dimensions of phase space provided in the TGAS catalog: positions, proper motions, and parallaxes ($\varpi$). Our Bayesian classification algorithm identified pairs of stars as being a chance coincidence or a wide binary based on a posterior probability that included likelihoods for both outcomes and data-driven priors. Importantly, to estimate contamination without any assumptions or preconceptions on Galactic structure, we repeated our search by matching the TGAS catalog with a shifted version of itself; every pair identified in this manner was a chance coincidence. A comparison between our wide binary candidates and our sample of chance coincidences provided a fully independent estimation of our contamination rate. For the sample generated with a power-law prior on the separation of wide binaries, having a posterior probability above 99\%, and with a projected orbital separation $s$~$<$~4$\times$10$^4$ AU\footnote{We additionally tested a log-flat prior, but found the power-law prior more representative of the data.}, we estimated a contamination rate of 6\%.  

\begin{figure*}
\begin{center}
\includegraphics[width=\textwidth,angle=0]{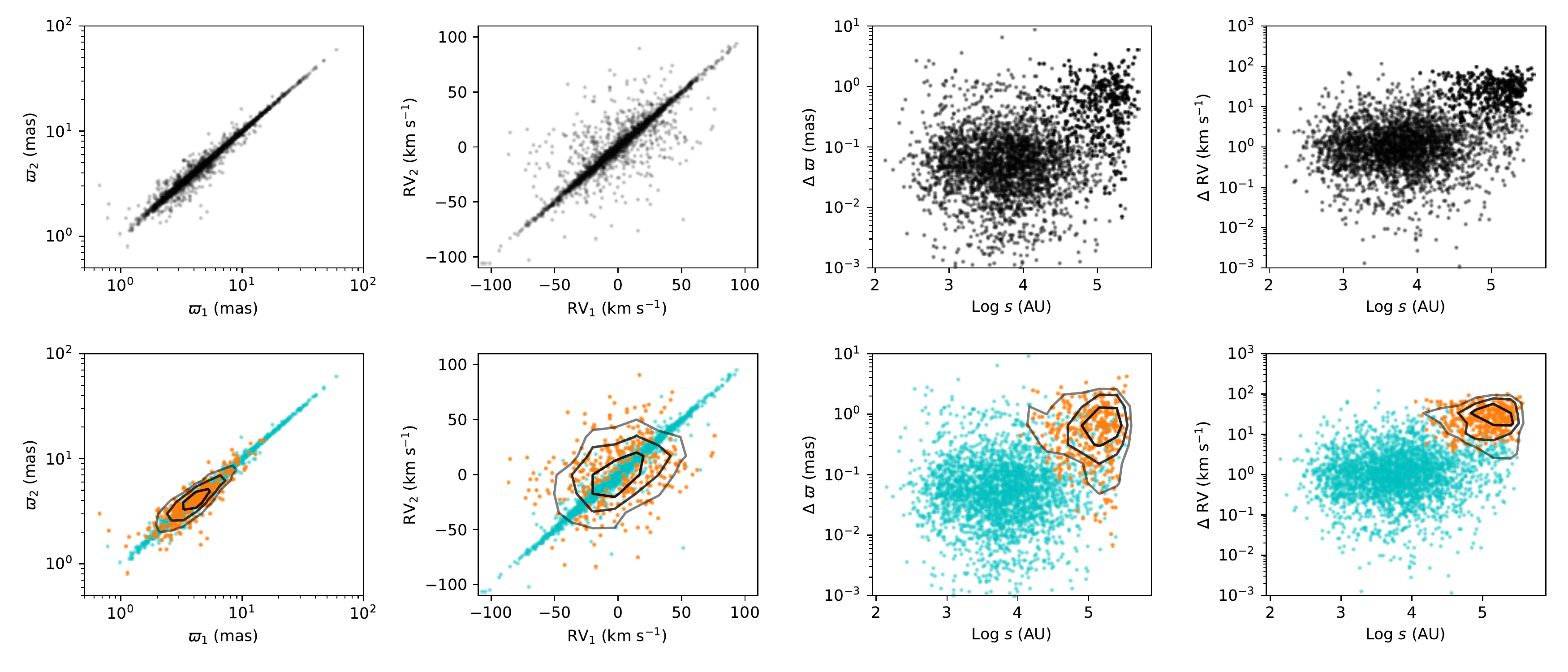}
\caption{ We compare the DR2 parallaxes (first column of panels) and RVs (second column of panels) for stars in the \citet{2017MNRAS.472..675A} TGAS wide binary sample. The right two columns of panels show the differences in these two values as a function of projected orbital separation. Points in the bottom panels are identical to those in the top panels, with colors indicating their association with one of two clusters as determined by an agglomerative clustering algorithm \citep{sklearn}. Black contours indicate the locus of our independently derived TGAS catalog of chance coincidences. The congruence between the black contours and the orange cluster suggests that candidate binaries associated with this cluster are predominantly contaminating pairs. \label{fig:1}}
\end{center}
\end{figure*}

To obtain an improved estimate of the contamination of our sample, we first cross-match our stars with the subset of the {\it Gaia} DR2 catalog \citep{2018arXiv180409365G} with RVs by selecting those stars with DR2 positions that are within 0\farcs1 of the TGAS positions. Of our 5409 binaries, 2992 (55\%) have {\it Gaia} DR2 counterparts with parallaxes and RVs for both components. In the top row, the left two panels of Figure~\ref{fig:1} show the parallaxes and RVs of the components of our wide binaries, and the right two panels show how the parallax and RV differences scale with $s$. 

While Figure~\ref{fig:1} shows a striking consistency between the parallaxes and RVs for most of our pairs, it does appear that there is a distinct population of outliers. To test this, we apply an agglomerative clustering algorithm \citep{sklearn} on the three dimensional log $s$-log $\Delta\varpi$-log $\Delta$RV data for the 2992 binaries in this matched TGAS/DR2 catalog. We find that the binaries do in fact separate into two samples. The bottom row of panels shows a reproduction of the top row, in which the color of the points indicates the membership assigned by the clustering algorithm to each pair. 

We then cross-match our separate sample of chance coincidences from TGAS with DR2. The black contours in the bottom panels of Figure~\ref{fig:1} show the positions of these chance coincidences; they clearly overlap with the systems at large $s$. Because of this congruence, we conclude that the orange set of points is comprised principally of chance coincidences, whereas the cyan points are genuine wide binaries. This strengthens our conclusions from \citet{2017MNRAS.472..675A} and \citet{2018MNRAS.473.5393A} that TGAS pairs at $\sim$pc scale separations and larger are mostly comprised of chance coincidences.

\section{}

Comparing the relative size of the two sets of pairs provides a contamination estimate of 13\% for the entire TGAS/DR2 sample and 2\% for the sample with $s$$<$$4\times10^4$ AU. These rates are somewhat underestimated, as the third column of panels shows a number of pairs with $\Delta \varpi$ $\approx$ 1, but with log $s$$<$4.5, which are likely contaminating pairs unidentified by the clustering algorithm. Including stellar pairs with $\Delta \varpi$/$\sigma_{\Delta \varpi}$$>$5 increases our estimated contamination rates to 16\% and 6\% for wide binaries over the entire range in $s$ and those with $s$$<$$4\times10^4$, respectively, rates that are in agreement with our initial estimate in \citet{2017MNRAS.472..675A}. 

From this exercise, we conclude that our sample of chance coincidences generated from a shifted version of our catalog provides an accurate estimate of the contamination rate in our sample. We confirm the high fidelity of our wide binary sample, justifying its use for investigating a host of astrophysical questions, such as the metallicity consistency of co-eval stars \citep{2018MNRAS.473.5393A}, testing gravity in the low acceleration regime \citep{2017arXiv171110867P}, and stellar multiplicity in moving groups \citep{2016MNRAS.459.4499E}. Furthermore, the low contamination rate of our TGAS sample suggests that our algorithm, or some form of it, can be analogously applied to accurately classify wide binaries among the $10^9$ stars with measured positions, proper motions, and parallaxes in {\it Gaia} DR2.

\software{scikit-learn \citep{sklearn}}

\end{document}